\documentclass{nature}
\usepackage{xcolor}
\usepackage{url}
\usepackage{amssymb,amsmath}
\newcommand{\TXS}{TXS\,0506+056 } 
\newcommand{\HE}{high energy neutrinos~}

\usepackage[locale=US, exponent-product=\cdot]{siunitx}
\usepackage[symbol]{footmisc}
\usepackage{todonotes}

\title{The Pacific Ocean Neutrino Experiment}

\usepackage{lineno}

\author{M. Agostini$^{6}$, M. B\"ohmer$^{6}$, J. Bosma$^{10}$, K. Clark$^{9}$,
M. Danninger$^{10}$,  
C. Fruck$^{6}$, R. Gernh\"auser$^{6}$, A. G\"artner$^{3,6}$, D. Grant$^{2}$, F. Henningsen$^{6,8}$,
K. Holzapfel$^{6}$, M. Huber$^{6}$, R. Jenkyns$^{11}$, 
C. B. Krauss$^{3}$, 
K. Krings$^{6}$, C. Kopper$^{2}$,
K. Leism\"uller$^{6}$,
S. Leys$^{4}$,
P. Macoun$^{11}$,
S. Meighen-Berger$^{6}$,
J. Michel$^{5}$, R.W. Moore$^{3}$,
M. Morley$^{11}$,
P. Padovani$^{7}$,
T. Pollmann$^{6}$,
L. Papp$^{6}$,
B. Pirenne$^{11}$,
C. Qiu$^{11}$,
I. C. Rea$^{6}$,
E. Resconi$^{6}$,
A. Round$^{11}$,
A. Ruskey$^{11}$,
C. Spannfellner$^{6}$, M. Traxler$^{1,4}$, A. Turcati$^{6}$, J.\,P. Yanez$^{3}$.}

\usepackage{graphicx}
\makeatletter
\let\saved@includegraphics\includegraphics
\AtBeginDocument{\let\includegraphics\saved@includegraphics}
\renewenvironment*{figure}{\@float{figure}}{\end@float}
\makeatother

\begin{document}
\maketitle
\begin{affiliations}
\item Helmholtzzentrum f\"ur Schwerionenforschung (GSI) Planckstrasse 1, 64291 Darmstadt, Germany
\item Department of Physics and Astronomy, Michigan State University, East Lansing, MI 48824, USA
\item Department of Physics, University of Alberta, Edmonton, Alberta, Canada T6G 2E1
\item Department of Biological Sciences, University of Alberta, Edmonton, Alberta, Canada, T6G 2E9
\item Institut f\"ur Kernphysik, Goethe Universit\"at, 60438 Frankfurt, Germany
\item Department of Physics, Technical University of Munich, D-86748 Garching, Germany
\item European Southern Observatory, Karl-Schwarzschild-Str. 2, D-85748 Garching, Germany
\item Max-Planck-Insitut f\"ur Physik, D-80805 Munich, Germany
\item Department of Physics, Engineering Physics \& Astronomy, 64 Bader Lane, Queen's University, Kingston, ON, Canada, K7L 3N6
\item Department of Physics, Simon Fraser University, 8888 University Drive Burnaby, B.C. Canada, V5A 1S6
\item Ocean Networks Canada, University of Victoria, Victoria, British Columbia, Canada
\end{affiliations}


\pagebreak
{\bf The Pacific Ocean Neutrino Experiment (P-ONE) is a new initiative with a vision towards constructing a multi-cubic kilometre neutrino telescope, to expand our observable window of the Universe to highest energies, installed within the deep Pacific Ocean underwater infrastructure of Ocean Networks Canada.}\\
The Universe is opaque to very high energy photons, limiting the horizon of $\gamma$-ray astronomy above 100\,TeV primarily to our Galaxy. Neutrinos, nominally the ideal astrophysical messenger, allow for the exploration of the cosmos up to the highest energy frontier. 
Following the IceCube Neutrino Observatory's discovery of an astrophysical flux of neutrinos in 2013\cite{Aartsen:2013bka, Aartsen:2013jdh},  and the following link between these neutrinos and a $\gamma$-ray emitting blazar in 2017\cite{IceCube:2018cha}, a global effort has mobilized to establish dramatic improvements in the integral exposure to astrophysical neutrinos. Ocean Networks Canada (ONC), a unique oceanographic observatory, offers a new opportunity for the construction of a large volume neutrino telescope. Among the various ONC-powered nodes, the Cascadia Basin at a depth of 2660 meters has been selected to host the Pacific Ocean Neutrino Experiment (P-ONE, see \url{http://www.pacific-neutrino.org}).  P-ONE as a new telescope for TeV-PeV neutrinos will build on a highly modular deployment and maintenance approach.

Neutrinos being neutral and interacting only through nature's weak force provide a unique opportunity to investigate phenomena in the Universe that cannot be probed by photons due to limited penetration depth, energy or absorption on their long path to Earth. With neutrinos, the physics of particle acceleration near black holes, particle interaction above the PeV scale, and in general, the fundamental laws of nature under the most extreme energy and gravitational conditions may be explored\cite{Ahlers:2015lln}. 
Besides being a complementary astrophysical messenger to the photon, and providing a baseline for oscillation studies,  high energy neutrinos hold the potential to reveal the identity and nature of dark matter\cite{Ibarra:2018yxq}, provide insights into the search for axions using extreme blazars as targets\cite{Biteau:2020prb}, and constrain the presence of heavy neutral leptons predicted in many extensions of the Standard Model of particle physics that could be produced in our atmosphere\cite{Coloma:2019htx}.  

\begin{figure}
\includegraphics[width=1.0\textwidth]{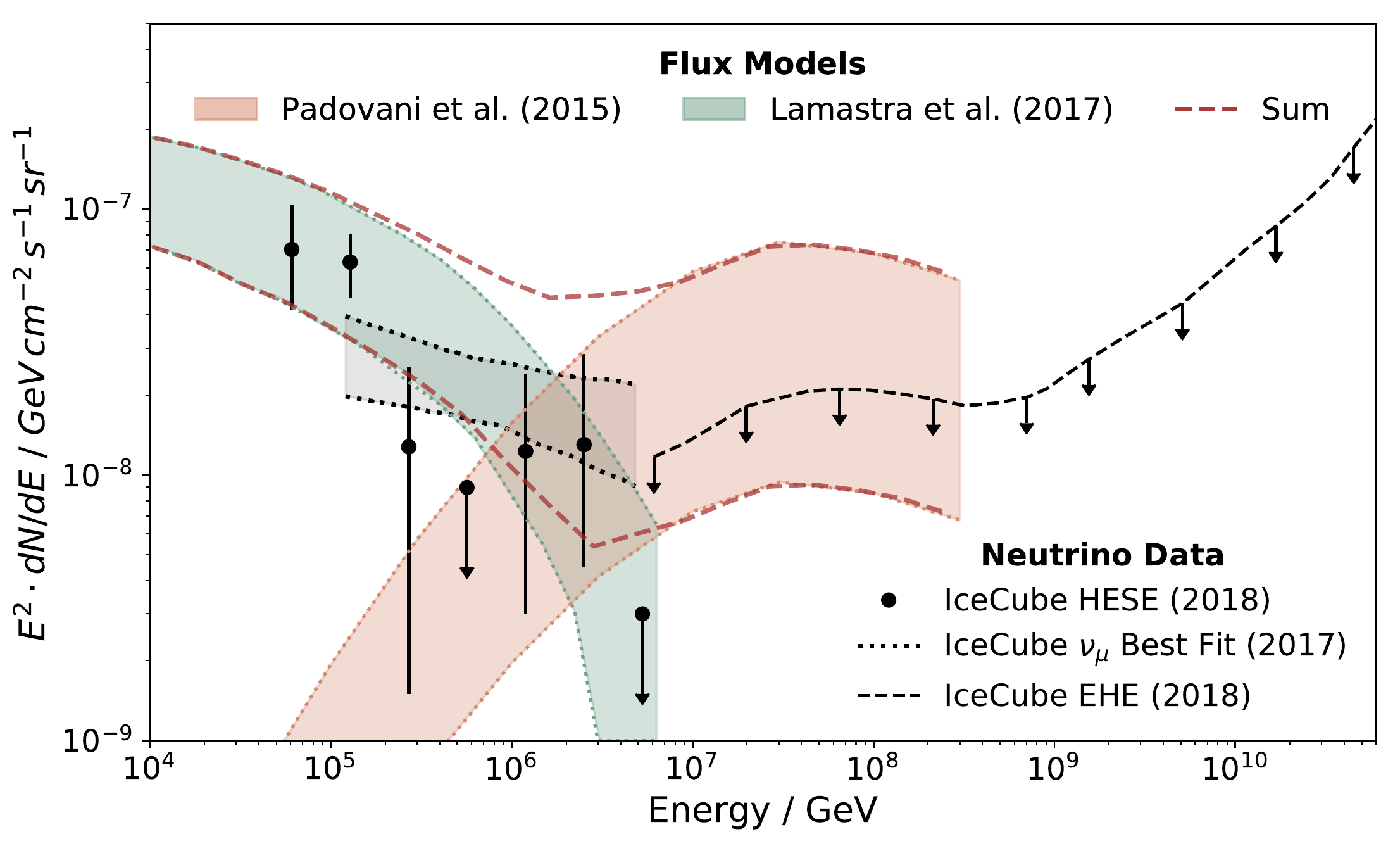}
\caption{Comparison between the diffuse flux of high energy neutrinos measured by IceCube (IceCube $\nu_\mu$ Best Fit, IceCube HESE, experimental points) and models related to neutrinos' contributions by 
AGN outflows\cite{Lamastra:2017iyo} (green band) and blazars\cite{Padovani:2015mba} (red band). The range in the latter indicates a neutrino-to-$\gamma$-ray flux ratio of 0.1 - 0.8. The dashed line represents the extreme high energy (EHE) upper limits of IceCube  (IceCube EHE)\cite{Aartsen:2018vtx}, which constrains this ratio.} 
\label{figure:Fig.1}
\end{figure}

To reach the sensitivity to detect high energy astrophysical neutrinos requires a detector volume of  several cubic-kilometers, as demonstrated by the IceCube South Pole neutrino telescope\cite{Aartsen:2013jdh, Aartsen:2016nxy} in its ten years of operation. 
IceCube, the only cubic-kilometer-scale neutrino telescope to date, has observed an isotropic signal of astrophysical neutrinos at the TeV-PeV scale\cite{Aartsen:2013bka, Aartsen:2013jdh, Aartsen:2014gkd,Aartsen:2017mau, Aartsen:2016xlq}. This groundbreaking discovery implies the existence of a yet unknown class of extragalactic astrophysical (non-thermal) objects, like starburst galaxies, active galactic nuclei (AGNs), and gamma-ray bursts (GRBs)\cite{Ahlers:2015lln}, accelerating protons up to at least 10$^{16}$--10$^{18}$\,eV.
Independent of the source class, production of high energy neutrinos requires some common astronomical ingredients: a proton-loaded jet, and; a sufficiently large reservoir of target photons/protons of the appropriate energy.
Realizing detectable neutrino sources from the combination of these ingredients
can be achieved in two extreme ways -  powerful but rare sources (e.g. blazars\cite{Padovani:2017zpf}), or weaker but more numerous objects (e.g. AGN outflows\cite{Lamastra:2017iyo, Padovani:2018hfm}), where reality most probably falls somewhere in between (see Fig.~\ref{figure:Fig.1}).
With eight years of monitoring the neutrino sky, IceCube obtained the first compelling evidence for an association between \HE and a known blazar\cite{IceCube:2018dnn, IceCube:2018cha, Padovani:2018acg}.
The evidence is based on a highly energetic neutrino-induced muon track (IceCube-170922A) that was followed up by the international astronomical community. A radio bright blazar, TXS\,0506+056 at $z=0.3365$\cite{Paiano:2018qeq}, was observed to be flaring in $\gamma$-rays  at a distance of 0.1 degrees from the direction of the neutrino event. The reconstructed angular uncertainty of the event was 0.4 degrees and the probability of a random neutrino-blazar coincidence is excluded at the 99.73\%\,C.L\cite{IceCube:2018dnn}. IceCube also independently found in its archival data an excess of neutrinos from the direction of the same blazar between October 2014 and February 2015. This excess, or flare, can be excluded to be of background origin at the level of 99.95\%\,C.L\cite{IceCube:2018cha}. {\it Fermi}-LAT data indicate a hardening of the spectrum during the same time period, pointing to a possible high energy (TeV) component\cite{Padovani:2018acg}, but no simultaneous data are available leaving uncertain the presence of a $\gamma$-ray activity in connection with the neutrino flare.\\
The diffuse neutrino flux measured by IceCube\cite{Aartsen:2016xlq} in the energy range  \SI{194}{\tera\electronvolt}--\SI{7.8}{\peta\electronvolt} is compatible with a single power-law component parametrized as:
\begin{align}
   \frac{\partial \phi_{\nu_{\mu}+\bar{\nu}_{\mu} }}{\partial E} = \phi_{\SI{100}{\tera\electronvolt}} \cdot  \left(\frac{E}{\SI{100}{\tera\electronvolt}}\right)^{-\gamma},  
\end{align}
with spectral index $\gamma=2.13\pm0.13$ and $\phi_{\SI{100}{\tera\electronvolt}} = \left(0.9^{+0.30}_{-0.27}\right)\cdot\SI{e-18}{\per\giga\electronvolt\per\square\centi\meter\per\second\per\steradian}$. 
The neutrino flare observed from the direction of \TXS corresponds to about 1\% of the entire integrated diffuse neutrino signal measured by IceCube. The observation of \TXS thus implies that the number of neutrino sources that can still be discovered is of the order of hundreds to thousands. In order to discover new neutrino sources, future neutrino telescopes will need to be able to explore the areas of $\phi_E < 10^{-12}$\,TeV\,cm$^{-2}$\,s$^{-1}$\cite{Aartsen:2019fau}.  
Where the neutrino flux produced by a single source is too weak for a significant association, methods such as stacking catalogued sources can be used to compensate for the lack of individual observations. However, the results obtained from stacking are only informative on the basis of specific assumptions about the brightness, spectral index and time pattern of the neutrino sources. For example, the results obtained by stacking the third catalog of hard {\it Fermi}-LAT sources (3FHL)\cite{TheFermi-LAT:2017pvy} show that the blazars in the 3FHL catalogue cannot explain the entire flux of diffuse astrophysical neutrinos\cite{Aartsen:2019fau}.
Moreover, their maximum contribution can vary significantly depending on the assumed spectral index of the source population\cite{Huber:2019lrm}.  
At present, the blazar scenario - anticipated by many authors even before the IceCube discovery\cite{Mannheim:1995mm, Halzen:1997hw} 
might explain a significant part of the IceCube neutrino signal above $\sim 0.5$ PeV, while  only contributing $\sim10\%$ at lower energies, leaving room  for some other population(s)/physical mechanism\cite{Padovani:2015mba, Padovani:2016wwn, Resconi:2016ggj}. 

\section*{Design of P-ONE}
\begin{figure}[h!]
\centering
\includegraphics[width=1.0\textwidth]{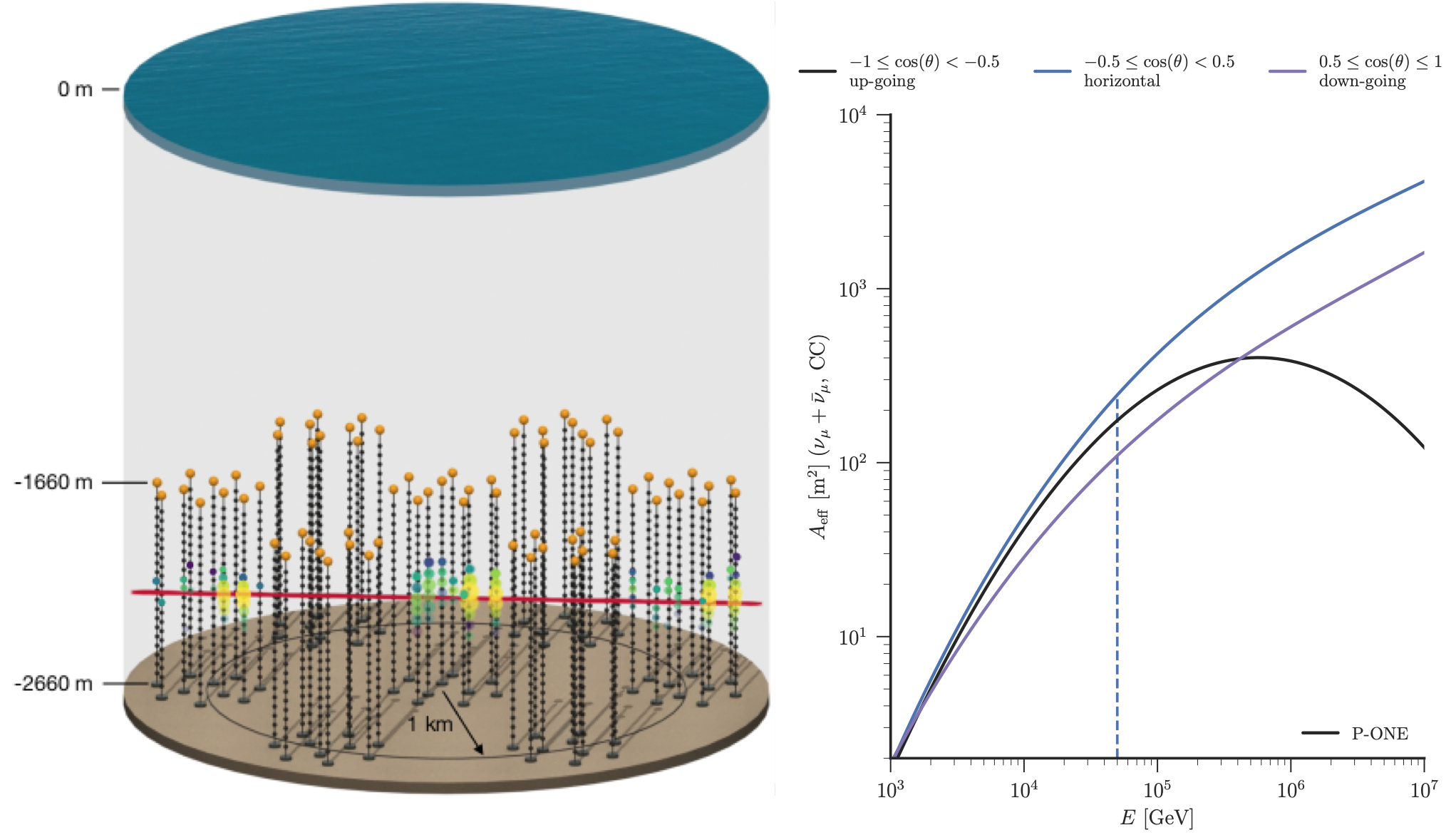}
\caption{(left) Design of the Pacific Ocean Neutrino Experiment, as presented in this work, consisting of seven sectors of 10 strings optimized for energies 10\,TeV-10\,PeV. The light pattern produced by a 50\,TeV neutrino is included shown within the detector. (Right) Neutrino and anti-neutrino effective areas for charged current interactions in P-ONE. The three different lines represent the effective area in different declination bands. The dotted line at 50\,TeV shows the equivalent energy of the event view in the detector. The effective area is calculated by requiring a minimum deposit of 100\,GeV of energy, and a minimum track of 100\,m in one sector.  The effective area calculations also take into account the absorption effect, which is more relevant for high-energy neutrinos transiting deep through the Earth\cite{Aartsen:2017kpd, Bustamante:2017xuy}. Neutrinos from the horizon are nearly absorption-free, even at energies above the PeV scale.}
\label{fig:Fig2}
\end{figure}

An era of mutli-cubic-kilometre neutrino telescopes is now technologically possible thanks to the experience gained over the last decade in the development of deep ocean infrastructure, for example by ONC, and in the construction of neutrino telescopes by IceCube\cite{Aartsen:2016nxy}, ANTARES\cite{Aguilar:2006rm} and Baikal-GVD\cite{Avrorin:2019dli}.  With the challenge of a stable system within which the construction and operation of deep ocean neutrino telescopes is largely addressed, the next hurdle is the physical structure of the  telescope itself and the associated costs. With a characteristic deep ocean (lake) water optical attenuation length of 20-70\,m, a uniform infill array, similar to IceCube\cite{Aartsen_2017}, would require thousands of instrumented lines distributed
over the entire volume. This poses a daunting challenge that,  following from the pioneering effort of the DUMAND project\cite{Babson:1989yy}, can be overcome by introducing a segmented neutrino telescope design (see Fig.\,\ref{fig:Fig2}).
For the search of individual astrophysical sources, P-ONE will primarily target horizontal high energy muon tracks, but will also be sensitive to high energy induced showers. 
 The muon produced in the interaction of a very high-energy neutrino can travel several kilometers in water or ice. The Cherenkov light produced by the propagating muon provides enough information for the reconstruction of the incoming neutrinos with an angular resolution of a fraction of a degree\cite{Aartsen_2017}. 
A segmented array sampling approach would retain sufficient information to reconstruct the incoming direction and the energy of the track while significantly reducing the number of mooring lines required to cover the large volumes  (see Fig.\,\ref{fig:Fig2}). 
The P-ONE {\it Explorer}, corresponding to the first 10-string segment, has a planned deployment in 2023-2024 during a marine operation scheduled to last four weeks. A total of 20 photo-sensors and at least two calibration modules will compose each individual string of the Explorer. The remainder of the telescope array, comprising on order 70 strings, are planned for deployment between 2028-2030. 
As the pathfinder stage for the complete telescope, the Explorer will not only pave the way for a successful P-ONE installation  but also improve on existing searches for $10-100$\,TeV energetic $\nu_\tau$ of astrophysical origin\cite{Aartsen:2020aqd}, probe the production of long-lived exotic particles in the atmosphere, and complete current studies on the systematic uncertainties of cosmic ray interaction patterns. 

P-ONE will be installed and operated within one of the world's largest and most advanced cabled ocean observatories, ONC. ONC is composed of many infrastructures, the largest being the North East Pacific Time-series Underwater Networked Experiment (NEPTUNE). 
The NEPTUNE cabled ocean observatory\cite{Neptune:2010},
completed in 2009,  comprises an 800\,km loop of telecommunication fibre-optic cables to power and transfer data to a variety of sensors connected to five nodes across the Juan de Fuca tectonic plate, approximately 200\,km off the coast of British Columbia, Canada.
The installation of the scientific instrumentation is effectively {\it plug and play}, allowing for highly modular deployment and maintenance. 

The node deployed in 2009 at Cascadia Basin at a depth of 2660\,m has been selected to host the P-ONE detector array. The wide and flat sediment surface provides ideal conditions to host a neutrino observatory, namely a large abyssal plain with temperature below 2\,$^o$C.  The Cascadia Basin is home to an assortment of well-adapted organisms, some of which emit bioluminescence light that is a source of background for a neutrino telescope. 
To fully qualify the Cascadia Basin as a site to host a multi-cubic-kilometer-scale neutrino telescope, a two-line calibration setup - the STRings for Absorption length in Water (STRAW)\cite{Bedard:2018zml} - was deployed in Spring 2018 and connected to the existing NEPTUNE node. STRAW has been continuously monitoring the deep Pacific Ocean conditions: without the need of any repair or maintenance, STRAW works in near real-time with a duty cycle of 98\% since completion of the commissioning phase in January 2019. 
STRAW has provided a first {\it in-situ} measure of the effective attenuation length -- at a wavelength of 465\,nm -- of 35\,$\pm$\,5\,m as well as the monitoring of the local bioluminescence activity over a period of more than 12 months. 
 The photon rate observed in the 400-600 nm wavelength window shows periodicity of 12 and 24 hours connected to the semi-diurnal and diurnal tides, respectively. The level of irradiance observed in the deep ocean is in the range of 10$^{-11}-10^{-9}$\,W/m$^2$, and hence quite low. No periods of significant increased environmental background were observed yet.
STRAW is scheduled to continue monitoring of the site through 2021.
 A second investigative mission, denoted STRAW-b, is presently under construction with deployment planned in 2020.  The STRAW-b advanced optical monitoring will continue the qualification of the site while, in parallel, the P-ONE collaboration finalizes the design of the {\it Explorer} detector array.

\section*{A distributed planetary network of telescopes}
From the first indicative high energy neutrino associations with blazars\cite{Giommi:2020hbx},
it is evident that neutrino astronomy finds itself at the tipping point. A global effort has been mobilized\footnote{\url{http://www.globalneutrinonetwork.org}} in order to fully explore the sky at the highest energies and to reveal the most powerful cosmic accelerators in the Universe. 
Contributing parties to the effort are the neutrino telescope ANTARES (see \url{https://antares.in2p3.fr}, in operation since 2008,  KM3NeT (see \url{https://www.km3net.org}), which is under construction in Italy and France, the Baikal-GVD (see \url{https://baikalgvd.jinr.ru}) detector under construction in Russia (Lake Baikal), 
IceCube and IceCube-Gen2 at the South Pole, and in the near future P-ONE. When combined to behave as a single distributed planetary instrument, the acceptance for astrophysical neutrinos at above the PeV-scale covers the entire sky. In fact, as recently verified\cite{Aartsen:2017kpd, Bustamante:2017xuy}, the cross section for the neutrino--nucleon interaction increases with increasing neutrino energy, making the Earth significantly less transparent to neutrinos with energies above 50\,TeV; less than 20\% (5\%) of the neutrinos with energy of 100 TeV (1\,PeV) can cross the Earth at vertical zenith angle cos($\theta$)\,=\,-0.8., see also Fig.\,\ref{fig:Fig2}. Neutrino telescopes are therefore effectively blind to very high energy neutrinos crossing the Earth.
In order to obtain peak all-sky neutrino exposure, it will be essential to combine the information from the various telescopes.
Each telescope is designed to send alerts to  the entire astronomy community, including the other neutrino telescopes, making it possible  to follow in real-time the temporal evolution of a transient event emitting in the PeV energy region (e.g. GRBs).
Therefore, while the P-ONE telescope is  designed to operate individually, providing partial sky coverage, the real impact for the exploration of the non-thermal universe via neutrinos is anticipated through leveraging the strengths of the global suite of telescopes as a unified instrument~\cite{resconi_elisa_2019_3520454}.
To quantify an order of magnitude gain from the combination of neutrino telescopes distributed around the Earth, we present here the neutrino acceptance defined as $\int_{0}^{\infty} A_{eff}(\delta, E_\nu) \cdot E_\nu^{-\gamma}$\,dE$_\nu$, where $A_{eff}$ is the neutrino effective area and $E_\nu$ the energy of the neutrinos. The gain in acceptance is sensitive to the intrinsic spectrum of possible neutrino sources; for comparison we therefore fix the neutrino spectral signal to $E_\nu^{-2.5}$ ($E_\nu^{-2.0}$).  In the scenario where four IceCube-equivalent telescopes are in operation around the globe (Lake Baikal, Mediterranean Sea, Pacific Ocean, and South Pole) an effective gain of more than a factor 100 (20) with respect to the single IceCube operation would be obtained. 
 The main improvement is observed in the Southern hemisphere, where IceCube data are dominated by atmospheric muons. Even including the atmospheric muon and neutrino veto capabilities\cite{Schonert:2008is}, the corresponding regions of the sky affected by this background results in a minimal effective area for neutrinos. Additional neutrino telescopes covering these regions with acceptance near the horizon will produce a significant relative increase in sensitivity.

In the search for persistent sources, the telescopes' data may be combined offline. If also sychronized in time, the network of neutrino telescopes provides a nearly all-sky exposure to high energy transient sources to the multimessenger community (e.g. Open Universe, \url{http://www.openuniverse.asi.it}), as well as the prompt follow up of alerts issued by gravitational wave detectors and gamma-ray facilities.  
An all-sky acceptance for neutrinos from the GeV to the PeV-scales in connection to the whole of multimessenger astronomy creates ideal conditions to discover the most extreme particle accelerators in the cosmos — the sources of the ultra high energy cosmic rays - and study the deepest mysteries of the non-thermal Universe.


\bibliographystyle{naturemag}
\bibliography{main}

\begin{addendum}
 \item We thank Stefan Sch\"onert, Daniele Vivolo, Philipp Eller for fruitful discussions. The authors are grateful and appreciative of the support provided by Ocean Networks Canada, an initiative of the University of Victoria funded in part by the Canada Foundation for Innovation.  This work is supported by the German Research Foundation through grant SFB\,1258 ``Neutrinos and Dark Matter in Astro- and Particle Physics'' and the cluster of excellence ``Origin and Structure of the Universe''. Support for Grant and Kopper is provided by Michigan State University.
 \item[Competing Interests] The authors declare that they have no
competing financial interests.
 \item[Correspondence] Correspondence and requests for materials
should be addressed to Elisa Resconi.~(email: elisa.resconi@tum.de).
\end{addendum}

\end{document}